\documentclass[conference]{IEEEtran}
\usepackage{graphicx}
\usepackage{shortcuts}
\usepackage{xspace}
\usepackage{code}
\usepackage{cite}
\usepackage{subcaption}
\usepackage{footnote}
\usepackage{mathtools}
\usepackage{url}
\usepackage{chngcntr}
\usepackage{listings, color, caption}
\usepackage{threeparttable}
\usepackage{setspace}
\usepackage{multirow}
\usepackage{multicol}
\usepackage{pifont}
\usepackage{soul}
\usepackage{boxedminipage}
\usepackage{booktabs}
\usepackage{multirow}
\usepackage{hyperref}

\usepackage{algorithm,algorithmicx}
\usepackage[noend]{algpseudocode}
\usepackage{indentfirst}
\usepackage{comment}
\graphicspath{ {images/} }
\usepackage{amsfonts}
\usepackage{mathrsfs}
\usepackage{amssymb}
\usepackage{amsthm}

\newcommand{\RN}[1]{%
  \textup{\uppercase\expandafter{\romannumeral#1}}%
}

\algrenewcommand\algorithmicindent{0.7em}%

\title{DeepVulSeeker: A Novel Vulnerability Identification Framework via Code Graph Structure and Pre-training
	Mechanism}

\author{
 \IEEEauthorblockN{Jin Wang\IEEEauthorrefmark{2},  Hui Xiao\IEEEauthorrefmark{2}, Shuwen Zhong\IEEEauthorrefmark{2}, Yinhao Xiao \IEEEauthorrefmark{1}\IEEEauthorrefmark{2}}

 \IEEEauthorblockA{\IEEEauthorrefmark{2}School of Information Science, Guangdong University of Finance and Economics, Guangzhou, China.}

 \IEEEauthorblockA{\IEEEauthorrefmark{1}Corresponding Author}

}

\begin{document}
\maketitle
\begin{abstract}
%Whether in public vulnerability reports or proprietary code, more and more software vulnerabilities are discovered every year. These vulnerabilities are likely to pose severe risks and even may give rise to system crashes and information leakage. As we know, people's lives are inevitably linked to the application of software, and therefore, software with security risks have a great impact on people's normal life, thus the identification and repair of vulnerabilities are utmostly significant for us. This is one item that has challenging work very much, we utilize the existing rich c open source code as well as the Deep Learning method to develop a vulnerability detection system, and we propose a model based on a heterogeneous graph neural network (MetaPath) and graph self-coding attention network (DeepVulSeeker), It is of great help to enhance the performance of parsing code semantic structure and node type structure. In addition, our model makes use of the method of combining four information structures——AST, CFG, DFG, and PLS to obtain more source code characteristics. Our evaluation of a wide range of data sets shows that DeepVulSeeker's performance is superior to the current mainstream models.

Software vulnerabilities can pose severe harms to a computing system. They can lead to system crash, privacy leakage, or even physical damage. Correctly identifying vulnerabilities among enormous software codes in a timely manner is so far the essential prerequisite to patch them. Unfortantely, the current vulnerability identification methods, either the classic ones or the deep-learning-based ones, have several critical drawbacks, making them unable to meet the present-day demands put forward by the software industry. To overcome the drawbacks, in this paper, we propose DeepVulSeeker, a novel fully automated vulnerability identification framework, which leverages both code graph structures and the semantic features with the help of the recently advanced Graph Representation Self-Attention and pre-training mechanisms. Our experiments show that DeepVulSeeker not only reaches an accuracy as high as 0.99 on traditional CWE datasets, but also outperforms all other exisiting methods on two highly-complicated datasets. We also testified DeepVulSeeker based on three case studies, and found that DeepVulSeeker is able to understand the implications of the vulnerbilities. We have fully implemented DeepVulSeeker and open-sourced it for future follow-up research.
\end{abstract}

\section{Introduction}
\label{sec:introduction}
Software vulnerabilities have always been a major threat in the software industry for a long time. They are still the main cause of the fragility of a computing system. It is reported by the U.S. department of commerce national institute of standards and technology (NIST) the national vulnerability database (NVD) that the number of vulnerabilities in 2021 reached 18,378, which is an unprecedented high record~\footnote{https://nvd.nist.gov/general/visualizations/vulnerability-visualizations/cvss-severity-distribution-over-time}. If they cannot be identified and fixed in a timely manner, they can pose serious potential threats to our daily life, e.g., possible financial theft from online shopping, private information leakage, or even physical damage~\cite{xiao2019brain}. Thus, swiftly and precisely discovering software vulnerabilities and fixing them is the key to guaranteeing the security of a computing system.

In academia, the research of vulnerability identification has been continually studied for years. In an early-stage study, vulnerability identification methods can be divided into three categories: static analysis~\cite{10.1007/978-3-642-16558-0_44}, dynamic analysis~\cite{10.1145/3106237.3106295}, and primitive machine learning methods~\cite{10.1145/1315245.1315311}. Static analysis refers to examining a software program, either with manual work or with the help of some forms of automated algorithms, without actually executing it. The main drawback of static analysis is that it either requires plenty of manual efforts or consumes a lot of time due to flow analysis~\cite{ming2015taint}. Different from static analysis, dynamic analysis requires actually running the program to locate bugs. Hence, dynamic analysis is not a scalable method due to its nature. Primitive machine learning methods refer to those that utilize basic machine learning schemes to identify vulnerabilities with pre-determined features. The weakness of these methods is that they have poor performance on complicated program datasets where features can vary drastically. 

In recent studies, the researchers tend to address the aforementioned shortcomings by applying Deep Learning, which is fully automated and is capable of being adapted to complicated situations. Some preliminary studies have shown that Deep Learning is a feasible method to identify more vulnerabilities than aforementioned methods~\cite{lineVD,2022arXiv220912181T,2022arXiv220910414N}. 
Yet, after we probe into these existing deep-learning-based works, we found that their performances are still under satisfaction due to several reasons. Some of the models, e.g., models based on pure Graph neural network (GNN), can only capture the structural features of a program~\cite{FUNDED}. The drawback of these models is that they neglect the rich semantic features (such as the function names and system calls) hidden within the codes. Some of the models merely extract the semantic features while ignoring the structural information presented in the codes~\cite{feng2020codebert}. Some of the works managed to consider the structural and semantic features, but their models were insufficiently trained for the task since they failed to realize the importance of the pre-training techniques~\cite{zhou2019devign}.

%It learns the vulnerability codes provided in vulnerability datasets to increase the accuracy of its judgment of existing or potential vulnerabilities. When exploring many vulnerability identification projects related to Deep Learning, we tried many advanced deep learning technologies, such as graph neural network and pre-trained models. Graph neural network (GNN) can capture the feature of the graph. When taking AST as a backbone, it can capture a large amount of a code snippet structure information. Without pre-trained mechanism, GNN loses the semantic information of a code snippet, resulting in the lack of shallow representation information of a code snippet in the model. Pre-trained models can accurately identify the semantic information of the code fragment, but they lack the structural information of a code snippet (such as AST, DFG and CFG). But they only obtain the surface semantics of the code snippet, losing the deep semantics of the code snippet.

To resolve the above drawbacks, in this paper, we propose DeepVulSeeker, a model based on a heterogeneous structural information with MetaPaths, pre-trained semantic models, and Graph Representation Self-Attention (GRSA) Encoding network. DeepVulSeeker first captures the semantic information relationship between each programmatic node and its neighbors of a piece of code with the help of a multi-layer bidirectional Transformer based pre-trained model named UniXcoder model~\cite{guo2022unixcoder}. UniXcoder can effectively encode the natural languages to program languages, providing more pertinent and precise semantic information for the training of our model. DeepVulSeeker then utilizes the semantic information to obtain the structural representations of the codes through Data Flow Graph (DFG), Control Flow Graph (CFG), and Abstract Syntax Tree (AST). In the end, DeepVulSeeker combines the obtained semantic information and the structural information, and feeds them to a convolution network and a feed-forward network to predict whether a given code snippet is vulnerable. In a nutshell, DeepVulSeeker can not only learn the structural information of a program, but also can understand its semantic implications. We fully implemented DeepVulSeeker and open-sourced the implementation on Github~\footnote{https://github.com/WJ-8/DeepVulSeeker}. We then evaluated DeepVulSeeker on standard Common Weakness Enumeration (CWE) datasets and two highly-complicated program datasets, i.e., QEMU and FFMPEG, with a total number of 18,519 pieces of codes, 8,125 of them are vulnerable, and the rest are bug-free. Our results show that DeepVulSeeker achieves a satisfactory performance on CWE datasets, with an accuracy as high as 0.99. As for the QEMU and FFMPEG datasets, DeepVulSeeker also has an acceptable outcome with an accuracy as high as 0.64, which, to the best of our knowledge, is higher than all existing well-known methods. We also conduct case studies demonstrating three typical cases that are not identified by other baseline approaches but were successfully identified by DeepVulSeeker. We further manually patched the vulnerabilities and fed them to DeepVulSeeker, and found out that DeepVulSeeker no longer flagged them as vulnerable, meaning that DeepVulSeeker is able to discover more complicated bugs and understand their meanings.
Our contributions are threefold:
\begin{itemize}
	\item We propose a novel vulnerability identification framework, namely DeepVulSeeker, which is capable of learning both the semantic information and the structural information of a given program, to determine if it is vulnerable. DeepVulSeeker can capture more proliferated features than other existing approaches in an automated manner. We fully implemented DeepVulSeeker and open-sourced it on GitHub.

	\item We conducted case studies that are absent in most of other research. The results showed that DeepVulSeeker is capable of finding more complicated vulnerabilities and better understanding their implications.
	
	\item We conducted plentiful experiments on large datasets and compared the performance with the most recent well-known methods. The results show that DeepVulSeeker outperformed most of these methods on the traditional CWE datasets and outperformed all methods on the highly-complicated program datasets, i.e., QEMU and FFMPEG.
\end{itemize}

\textbf{Paper Organization.} 
The rest of the paper is organized as follows. Section~\ref{sec:background} presents the recently advanced background knowledge of our approach. Section~\ref{sec:approach} details the design and technical components of the DeepVulSeeker framework. Section~\ref{sec:implementation} demonstrates our implementation of DeepVulSeeker. Section~\ref{sec:eval} reports our evaluation results and case studies on DeepVulSeeker. Section~\ref{sec:related} outlines the most related work.  Section~\ref{sec:conclusion} concludes the paper with a future research discussion.
\section{Background}
\label{sec:background} 
In this section, we mainly demonstrate the background knowledge of some recently advanced technologies that are exploited by DeepVulSeeker.

\subsection{Pre-trained Models}
Pre-trained models are machine learning models that are trained, developed, and supplied by other developers. They are typically used to solve problems based on deep learning and are always trained on large datasets. large-scale pre-trained models (PTMs) such as BERT and GPT have recently achieved great success and have become milestones in the field of artificial intelligence (AI). The pre-training mechanism has been developed as a typical machine-learning paradigm for decades. Thanks to complex pre-training objectives and huge model parameters, large-scale PTM can effectively extract knowledge from large amounts of labeled and unlabelled data. As widely demonstrated by experimental validation and empirical analysis, the wealth of knowledge implicit in huge parameters can benefit a variety of downstream tasks by storing it in huge parameters and fine-tuning it for specific tasks. The consensus in the AI community is now to adopt PTM as the backbone for downstream tasks, rather than learning models from scratch.
With the development of deep neural networks in the NLP community, the introduction of Transformers has made it possible to train very deep neural models for NLP tasks. In 2018, PTM, such as GPT and BERT, was proposed for NLP tasks, using Transformers as the architecture and targeting language model learning. From GPT and BERT, we can see that when the size of PTM becomes large, with hundreds of millions of parameters, large-scale PTMs can capture multiple sense disambiguation, lexical and syntactic structure, and factual knowledge from text. The rich linguistic knowledge of PTMs brings excellent performance to downstream NLP tasks by fine-tuning the large-scale PTMs on a large sample size. Over the past few years, large-scale PTM has performed well on both language comprehension and language generation tasks, even achieving better results than humans. There is also a consensus to fine-tune large-scale PTMs for specific AI tasks, rather than learning models from scratch.

\subsection{Graph Neural Networks (GNN)}
A graph neural network (GNN) is an optimizable transformation of all the graph properties (nodes, edges, global context) that preserve graph symmetry. This transformation means that ordering the vertex in another way does not change the connectivity of the graphs, and would not affect their physical meaning. Graph neural networks can be divided into five categories: Graph Convolution Networks, Graph Attention Networks, Graph Autoencoders, Graph Generative Networks, and Graph Spatial-temporal Networks.
\subsubsection{Graph Convolution Networks (GCN)}
The core idea is to learn a functional map by which a node in the map can aggregate its characteristics with the characteristics of its neighbors to generate a new representation of the node. GCN methods can be divided into two categories: spectral-based and spatial-based.
\begin{itemize}
\item The spectral-based method introduces a filter from the perspective of graph signal processing to define graph convolution, where the graph convolution operation is interpreted as removing noise from the graph signal.
\item In the space-based method, the graph convolution is represented as the aggregation of feature information from the neighborhood. When the algorithm of the graph convolution network runs at the node level, the graph pooling module can be intellect with the graph convolution layer to coarse the graph into a high-level substructure.
\end{itemize}

\subsubsection{Graph Attention Networks (GAN)}
The attention mechanism has been widely applied to sequence-based tasks. Introducing the attention mechanism into GNN enables neural networks to focus on nodes and edges that are more relevant to tasks. It uses attention in the process of aggregation, integrates the output of multiple models, and generates random walks oriented to important targets, which can effectively improve the effectiveness of training and the accuracy of testing. This forms the graph attention network. The followings are some of the most popular uses of attention mechanisms:
\begin{itemize}
\item Graph Attention Network (GAT) is a space-based graph convolution network. Its attention mechanism is used to determine the weight of the node neighborhood when the feature information is aggregated.

\item Gated Attention Network (GaAN) uses a multi-head attention mechanism to update the hidden state of the nodes. It uses a convolutional subnetwork to control the importance of each attention head.

\item Graph Attention Model (GAM) provides a recurrent neural network model to solve the graph classification problem by adaptively accessing a sequence of important nodes to process graph information.
\end{itemize}

\subsubsection{Graph Autoencoders}
Graph autoencoder is a kind of graph embedding method, whose purpose is to utilize neural network structure to represent graph vertices as low dimensional vectors. The common solution is to obtain the node embedding by using the multi-layer perceptron as the encoder, in which the decoder reconstructs the neighborhood statistics of the node.

\subsubsection{Graph Generative Networks}
Graph Generative Network (GGN) is a kind of GNN used to generate graph data. It uses certain rules to recombine nodes and generative edges and ultimately generates target graphs with specific properties and requirements.

\subsubsection{Graph Spatial-temporal Networks}
The graph space-time network also captures the space-time correlation of the space-time graph. The space-time graph has a global graph structure, with the input of each node changing over time.

\subsection{Self-Attention Mechanism}
The attention mechanism refers to forcing a deep learning network to focus on a particular piece of information and reducing attention to other pieces of information during the training process. The introduction of the attention mechanism in a neural network can reasonably allocate resources, focus on the key and useful information in a considerable amount of input information, and reduce the attention to the rest of the information, which can improve the efficiency and accuracy of the processing task. Thus, it is an effective way to solve information overload.

In the experiment of our study, we utilize a self-attention mechanism, which is a variant of the attention mechanism. It differs from the attention mechanism in that the self-attention mechanism allows the modeling of the dependence relationship without considering the distance between input information and output information. It can be utilized to calculate the correlation attention mechanism between different positions in a single sentence, meaning that it can calculate the mutual influence of different words in a single sentence to achieve global correlation weight. Therefore, the self-attention mechanism is an appropriate technique for our model since it can analyze the correlation of different locations in the program for the input code snippets in order to gain a better understanding of the code structure.

\section{Design of DeepVulSeeker}
\label{sec:approach}

\begin{figure*}[htbp]
	\includegraphics[width=190mm]{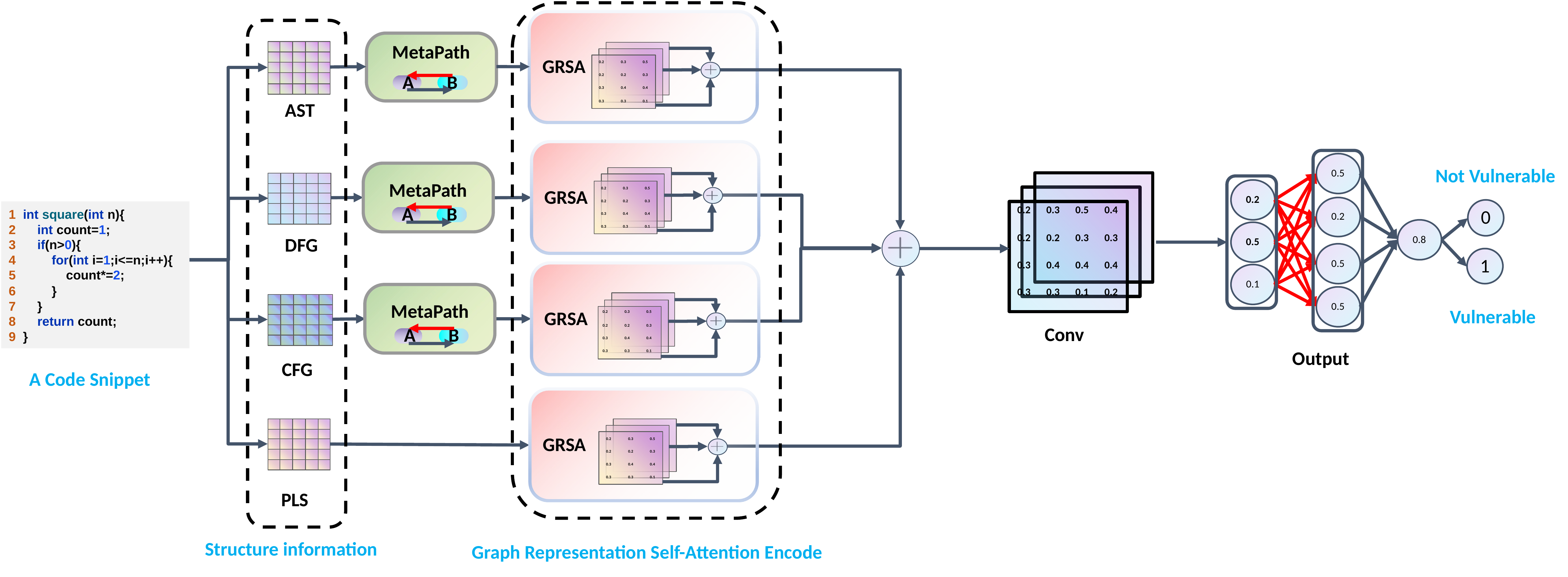}
	\centering
	\caption{Model Architecture}
	\label{fig:model}
\end{figure*}

In this section, we detail the design of DeepVulSeeker. The overall structure of DeepVulSeeker is shown in Fig.~\ref{fig:model}. 
The workflow of DeepVulSeeker can be described as a four-step process. First, DeepVulSeeker takes the target source codes as inputs and obtains the structural information, i.e., Data Flow Graph (DFG), Control Flow Graph (CFG), and Abstract Syntax Tree (AST) graphs, represented as adjacency matrices. Second, DeepVulSeeker encodes the raw source codes into contextualized token representations, i.e., Programming Languages Sequencing (PLS), empowered by the pre-training mechanism. Third, DeepVulSeeker further applies the technique of MetaPaths to the structural information and feeds them to Graph Representation Self-Attention Encode (GRSA) network together with the PLS outcome. Last, DeepVulSeeker forwards the outcomes from the third step to a convolution network and a feedforward network, which predicts if the given target source codes are vulnerable.

% DeepVulSeeker is mainly composed of three components: (1) Structure Information of source codes: We take source codes as inputs and encode a piece of raw source code into an adjacency matrix and contextualized token representations. (See in Sec.\ref{Sec.structure information}) (2) Multi-Metapaths: It captures the unique semantics between each node types pair and their neighbors and leaves longer MetaPaths for future evaluations in the adjacency matrix from structure information. (See in Sec. \ref{metapath}) (3) Graph Representation Self-Attention Encode: It takes structure information with Multi-Metapaths as a backbone and allows the model to jointly attend to information from different representations at different positions and captures more valid structure information. (See in Sec. \ref{grse}) 

\subsection{Structural Information}
\begin{figure}[ht]
	\includegraphics[width=60mm]{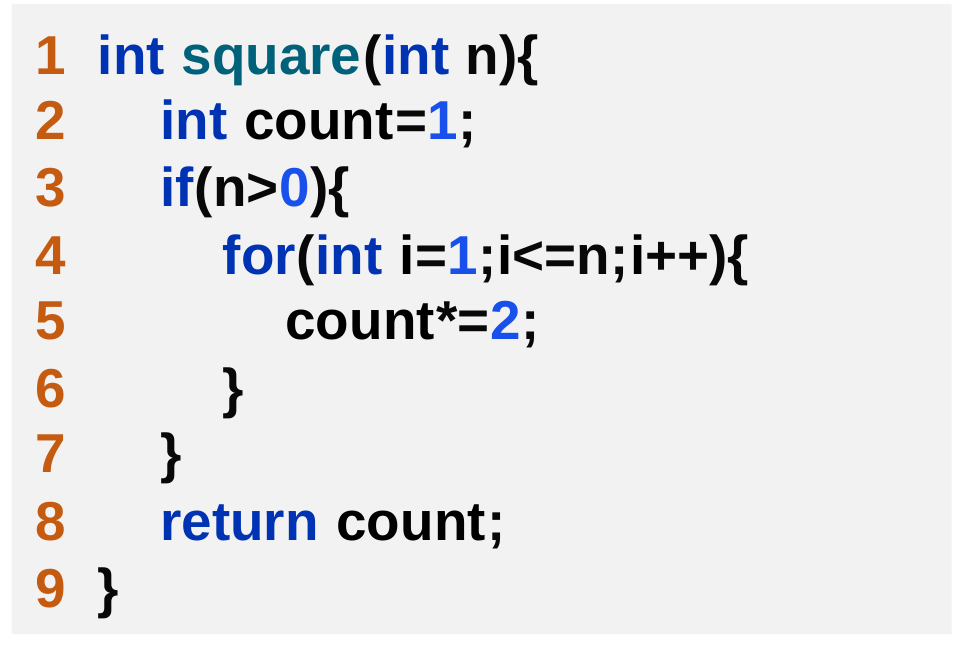}
	\centering
	\caption{A Code Example}
	\label{fig:example_code}
\end{figure}

\begin{figure*}[ht]
	\includegraphics[scale=0.45]{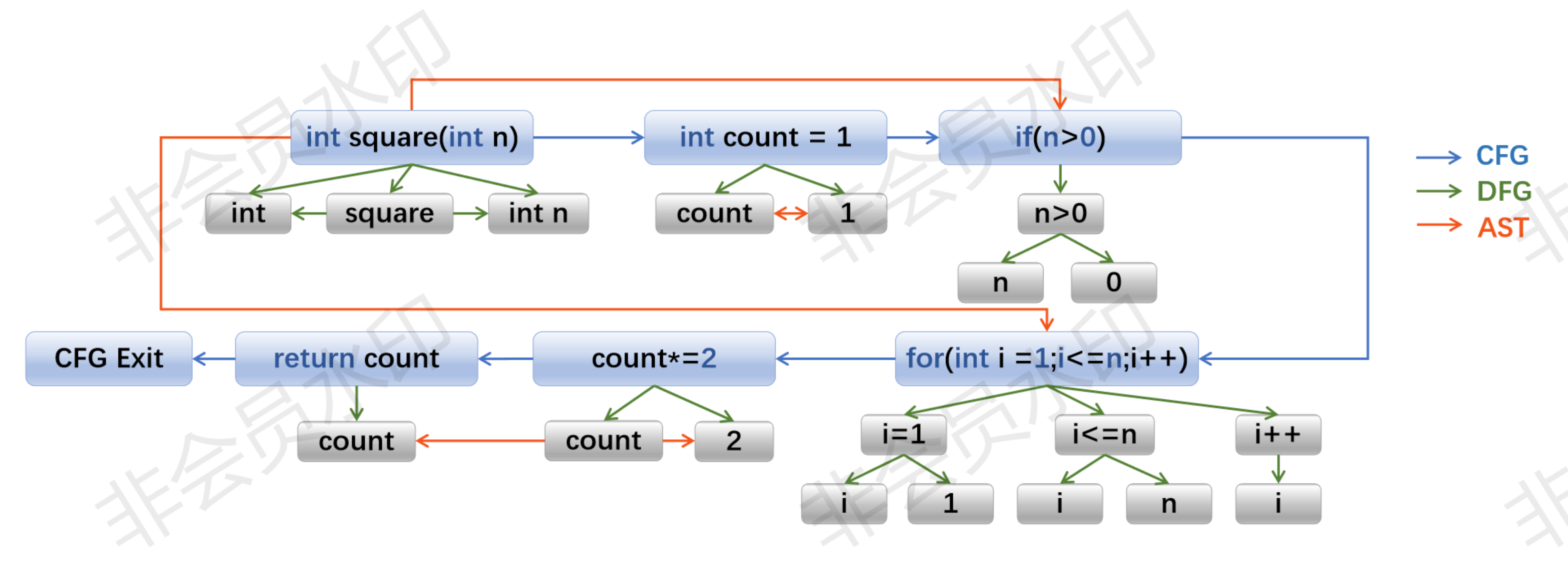}
	\caption{Generation of Structural Information of Fig.~\ref{fig:example_code}}
	\label{fig:ASTCFGDFG}
\end{figure*}

Structural information is one of the most critical parts of DeepVulSeeker. In this subsection, we leverage an example shown in Fig.~\ref{fig:example_code} and Fig.~\ref{fig:ASTCFGDFG} to illustrate how the structural information is generated.
\label{Sec.structure information}

\subsubsection{Abstract Syntax Tree (AST)}
AST is a tree representation of the abstract syntax structure of a piece of source codes. Each vertex in the tree represents a syntactic occurring of the code context, and each edge reflects the containment relationship between a pair of vertices. If we construct the AST for the example in Fig.~\ref{fig:example_code}, the result is a directed graph shown in Fig.~\ref{fig:ASTCFGDFG} in which the vertices are connected with the red edges. In this example, the token ``\texttt{count}'' in ``\texttt{int
count=1}'' is connected with ``\texttt{int}'', ``\texttt{count}'', ``\texttt{=}'' and ``\texttt{1}''.

\subsubsection{Control Flow Graph (CFG)}
CFG describes all possible control paths in a program that can be traversed. The paths to be traversed are specified by the branching statements, e.g., $ \mathit{if}$, $\mathit{while}$ and $\mathit{do}$ statements. In CFG, each vertex represents a syntactic occurring, and each edge stands for a possible branch. If we construct the CFG for the example in Fig.~\ref{fig:example_code}, the result is a directed graph shown in Fig.~\ref{fig:ASTCFGDFG} in which the vertices are connected with the blue edges. When the program hits the statement ``\texttt{if(n\textgreater0)}", it decides whether to run ``\texttt{for(int i=1; i\textless=n; i++)}'' based on \texttt{n} variable. Thus, the tokens in ``\texttt{if(n\textgreater0)}" are linked to the tokens within the "\texttt{if(n\textgreater0)}" statement and the tokens in ``\texttt{return count}".

\subsubsection{Data Flow Graph (DFG)}
DFG tracks the access of variables in a program. The access of a variable refers to both its value assignments and value modifications. In DFG, each vertex is a syntactic occurring, and each edge between a pair of vertices indicates that these two vertices are relevant to the assignments and modifications of a variable. If we construct the DFG for the example in Fig.~\ref{fig:example_code}, the result is a directed graph shown in Fig.~\ref{fig:ASTCFGDFG} in which the vertices are connected with the green edges. The variable "\texttt{count}", which is modified with a new value in the statement ``\texttt{count*=2}", is declared in the statement ``\texttt{int count=1}". Therefore, all the tokens in the statement of "\texttt{count*=2}" and those in "\texttt{int count=1}" are connected.

\subsection{MetaPath}
\label{metapath}
A MetaPath $\theta$ is a path in the form of $\theta = A_1\stackrel{R_1}{\longrightarrow}A_2\stackrel{R_2}{\longrightarrow} ... \stackrel{R_l}{\longrightarrow} A_{l+1}$, where $A_i$ is a state and $R_i$ is a composite relation between $A_i$ and $A_{i+1}$~\cite{liu2021smart}. The notation $\circ$ denotes the composition operator on relations. The length of a MetaPath is determined by the number of relations according to a specific model. In our model, $A_i$ refers to a vertex from the aforementioned graphs, and $R_i$ indicates a new edge that should be added between $A_i$ and $A_{i+1}$ based on a customized rule. In our design, DeepVulSeeker adopts a so-called length-2 MetaPaths rule~\cite{wang2019heterogeneous,sun2011pathsim}. According to this rule, if there is a directed edge between two vertices, we add a reverse directed edge between them. Most subgraphs from AST, CFG and DFG are tree-like, meaning that there are very few ``loops'' in the graphs. Training tree-like graphs may lead to the gradient vanishing problem. Therefore, we employ length-2 MetaPaths rule to mitigate this problem by improving the completeness of the graphs.

\subsection{Programming Languages Sequencing (PLS) }
PLS is the key process for DeepVulSeeker to understand the semantic implications behind the codes. It encodes each programmatic token into a vector of features, namely contextualized token representations, based on a pre-trained model. Compared with some traditional word embedding methods which require thorough training, PLS employs pre-training techniques to avoid the model being overfitting when the size of the training data is insufficient or skewed. Therefore, pre-training techniques offer more apposite features.
We denote $x_i$ as a given piece of codes, and the representations $P_i$ are obtained by Eq.~\ref{eq_pls}.
\begin{equation}
P_i= model(x_i) \label{eq_pls} ,
\end{equation}
where $model$ is the mathematical expression of a pre-trained model.

\subsection{Graph Representation Self-Attention Encoding}
\label{grse}
\begin{figure}[ht]
	\includegraphics[width=60mm]{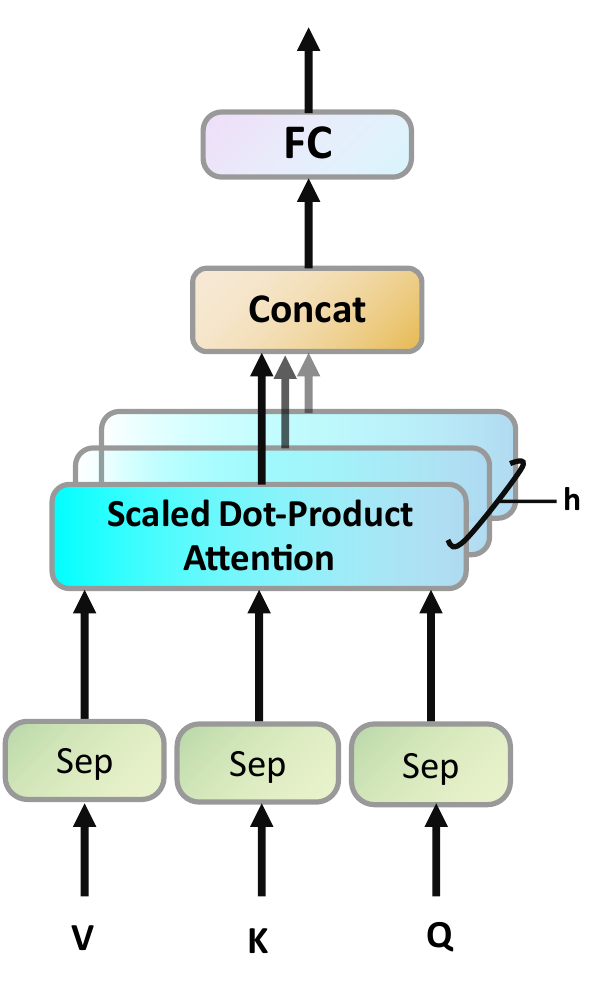}
	\centering
	\caption{The Structure of GRSA}
	\label{fig:multi-head}
\end{figure}

As aforementioned, we obtain three structural information (i.e., AST, CFG, and DFG), and semantic information from PLS. Our next step is to jointly combine these representations and feed them to our model for the follow-up training phase. In order to do so, we design the Graph Representation Self-Attention Encoding (GRSA) based on the multi-head self-attention mechanism~\cite{vaswani2017attention}. GRSA is a three-step process shown in Fig.~\ref{fig:multi-head}. First, GRSA takes three matrices, $Q$, $K$ and $V$, as inputs, and equally separates each matrix into multiple submatrices according to the number of heads (the selections of numbers are detailed in Sec~\ref{sec:implementation}). The notations of $Q$, $K$ and $V$ are query, key, and value matrix respectively in the attention-based model. Since we adopt a self-attention mechanism, $Q$, $K$ and $V$ are the same. In our model, we assign CFG, DFG, AST and PLS to $Q$, $K$, and $V$ each time. Second, GRSA retrieves the submatrices output by the first step and feeds them to a layer of scaled dot-product attention, in which the attention is calculated as Eq.~\ref{eq_selfatt_approach}.
\begin{equation}
\label{eq_selfatt_approach}
	h_i=\operatorname{softmax}\left(\frac{Q_i K_i^{T}}{\sqrt{d_{k}}}\right) V_i,
\end{equation}
where $Q_i$, $K_i$ and $V_i$ denote $i$-th submatrix of $Q$, $K$, and $V$, respectively. The notation of $d_k$ refers to the dimension of $K$. In the end, GRSA concatenates all the submatrices $h_i$ output from the scaled dot-product attention layer, and feeds them into a fully-connected network. After the process of GRSA, we manage to synthetically and effectively combine all information from the four different representations into one matrix, which contains both structural and semantic information.

% \begin{equation}
% \operatorname{GRSA}(Q, K, V)=\operatorname{Concat}\left(\operatorname{h}_{1}, \ldots, \operatorname{h}_{\mathrm{n}}\right) W^{O}\label{eq_multihead_approach},
% \end{equation}
% where $h_i$ correspond to the representation learned from the self-attention heads denoted by Eq.\ref{eq_selfatt_approach}, concat means the concatenation operation and $ n $ set via practice. In the end, we calculate all structure information and $H_{CFG}$, $H_{DFG}$, $H_{AST}$ and $H_{PLS} $ are obtained by GRSA, corresponding to the four structural information matrices.

\section{Implementation }
\label{sec:implementation}

In this section, we detail our implementation of DeepVulSeeker shown in Fig.~\ref{fig:model}.

\subsection{Generating Structual and Semenatic Information}

As demonstrated in Sec~\ref{sec:approach}, the first and second steps are to generate CFG, DFG, AST and PLS information. 

\subsubsection{Generating CFG and DFG Adjacency Matrices}
We extracted C++ code by using tree-sitter-c~\footnote{https://tree-sitter.github.io/tree-sitter/}, which is a parser generator tool and incremental parsing library that builds concrete syntax trees for source files and efficiently updates the syntax trees when editing the source files. After that, We built a Node class that records the header nodes, end node and the next node of the current node, so that later we can traversal all the nodes and their connected ones conveniently. We then used lists to store all the connected nodes, added directed edges among these nodes if they are dependent according to conditional statements, and generated the CFG adjacency matrix based on these directed edges. For example, we have a node 1, which has two child nodes 2 and 3. Then we set the matrix $M(1,2)=1$, $M(1,3)=1$. 
The generation process of the DFG adjacency matrix is similar to the one of CFG adjacency matrix. The only difference is that we added directed edges based on data flow instead of conditional statements.

\subsubsection{Generating AST Adjacency Matrix}
We utilized Joern~\footnote{https://github.com/joernio/joern}, an open-source source code analyzer, to parse the source codes to generate Code Property Graph (CPG), which contains all the edge information as the form of $i \rightarrow j$, indicating that there is a directed edge pointing from node $i$ to node $j$. We then generated the corresponding adjacency matrix according to the edge information.
 
\subsubsection{PLS Embedding}
We employed the HuggingFace \cite{wolf-etal-2020-transformers} Transformers framework, an integrated platform for sharing and loading pre-trained models, for PLS implementation~\cite{wolf-etal-2020-transformers}. However, we found that the pre-train models from HuggingFace incorrectly sliced the tokens for our datasets, by simply applying the slicing method for NLP to process the program codes. For example, the token \texttt{ff\_insert\_inpad}, which should not be split since it represents a variable in the code, was incorrectly split into tokens \texttt{ff}, \texttt{\_}, \texttt{insert}, \texttt{\_} and \texttt{inpad} by HuggingFace. During our practice, we discovered that incorrect slicing could greatly undermine the accuracy of our model. To resolve the problem, we utilized NLTK, an open-source natural language processing tool, to perform the token slicing task~\cite{bird2009natural}. By using this tool, we were able to correctly slice the tokens from all pieces of source codes, and fed the tokens a the UniXcoder pre-trained model, a unified cross-model pre-trained model for programming languages that supports us to understand code token sequences~\cite{guo2022unixcoder}. In the end, for every piece of source codes, UniXcoder outputs a PLS embedding matrix that contains its semantic implications. Note that UniXcoder can only accept a number of 512 tokens at maximum. Thus, we trimmed the tokens from those that have more than 512 tokens to meet the requirement of UniXcoder.

\subsection{Implementation of MetaPath and GRSA}

The third step of the DeepVulSeeker process involves with the implementation of MetaPath and GRSA.

\subsubsection{MetaPaths}
\label{imple:metapath}
As mentioned in Sec~\ref{sec:approach}, we leveraged the length-2 MetaPaths rule to realize the implementation of the MetaPaths. To do so, we wrote a Python program to add reversely-directed edges to all pairs of connected nodes. Using length-2 MetaPaths rule can mitigate the gradient vanishing problem without leading to path explosion problem as length-n MetaPaths rule does~\cite{wang2019heterogeneous, sun2011pathsim}.

\subsubsection{GRSA}
As mentioned in Sec~\ref{sec:approach}, our design of GRSA is shown as Fig.~\ref{fig:multi-head}. We implemented the design based on Keras~\cite{chollet2015keras}, an open-source Python-based deep learning library that is empowered by TensorFlow~\cite{abadi2016tensorflow}. In the end, the outputs of GRSA layer are fed to a convolution network and a feed-forward fully-connected network, which would be trained to predict whether a given code snippet is vulnerable. We employed the cross entropy loss function for the training of our model, which is calculated as follows:
\begin{equation}
	\mathcal{L}_{CE}=\sum_{i=0}^{N} y_{i} \log p_{i}+\left(1-y_{i}\right) \log \left(1-p_{i}\right)\label{eq_ce},
\end{equation}
where $y_i$ denotes the ground-truth labels and $p_i$ is the probability of the label generated by the model.

\section{Evaluation}
\label{sec:eval}

In this section, we detail the evaluation of DeepVulSeeker. We first demonstrate the experimental setup for the evaluation. We then describe the recently state-of-the-art baseline methods and the general comparison performance between these methods and our model. We also conduct the ablation study and case study to further explore and explain the potential of our model.

\begin{table*}[htbp]
	\centering
	\caption{Dataset Information}
	\label{table:dataset}
	\begin{tabular}{l|llll|ll}
		\hline
		\textbf{Project} & \textbf{Training Set} & \textbf{Validation Set} & \textbf{Test Set} & \textbf{Total} & \textbf{Vul} & \textbf{Non-Vul} \\ \hline
		\textbf{FFmpeg}  & 3958                  & 462                     & 499               & 4919           & 2438         & 2481             \\
		\textbf{QEMU}    & 10903                 & 1378                    & 1318              & 13600          & 5687         & 7913             \\
		\textbf{CWE-362}    & 451                 & 56                   & 57              & 564          & 189         & 375             \\
		\textbf{CWE-476}    & 1298                 & 162                   & 163              & 1623          & 396         & 1227             \\
		\textbf{CWE-754}    & 4034                 & 504                    & 505              & 5043          & 1359         & 3684             \\
		\textbf{CWE-758}    & 1089                 & 136                    & 137              & 1362          & 367         & 995             \\ \hline
	\end{tabular}
\end{table*}

\subsection{Experimental Setup}
We mainly demonstrate the performance metrics, datasets, and environment configuration used for evaluation.

\subsubsection{Performance Metrics}
we use two mostly used higher-is-better metrics in learning-related studies, i.e., accuracy and F1, for evaluation. We give the definitions of these two metrics in the following context. Note that we need to explain the definitions of precision and recall before explaining the one of F1.

\textbf{Accuracy:} The ratio of correctly labeled cases to the total number of test cases.\par
\textbf{Precision:} The ratio of correctly predicted samples to the total number of samples that are predicted to have a specific label. This metric answers questions such as ``Of all the code revisions that are labeled to be vulnerability-relevant, how many are correct?''. High Precision indicates a low false-positive rate.\par
\textbf{Recall:} The ratio of correctly predicted samples to the total number of test samples that belong to a class. This metric answers questions such as ``Of all the vulnerable test samples, how many are labeled to be vulnerable?''. High recall suggests a low false-negative rate.\par
\textbf{F1 Score:} The harmonic mean of Precision and Recall. It is an important performance indicator of a model when the test data have an uneven distribution of vulnerability types. F1 is calculated as follows:
$$2 \times \frac{\text { Recall } \times \text { Precision }}{\text { Recall }+\text { Precision }}$$\par

\subsubsection{Dataset}
We evaluate our approach on six C/C++-language datasets that are used in many state-of-the-art studies~\cite{zhou2019devign, wang2021codet5, wang2020combining}. The dataset contains four conventional software assurance reference datasets, namely, CWE-362, CWE-476, CWE-754, and CWE-758~\cite{NISTSoft48:online}. These four datasets are collected from various software that have relatively simple code structures and functionalities. Thus, their vulnerabilities may be easier to be identified. The other two datasets are FFmpeg and QEMU. FFmpeg is powerful a cross-platform open-source solution to record, convert and stream video~\cite{tomar2006converting}. QEMU is an extremely powerful operating-system-emulator that allows running operating systems for any machine, on any supported architecture~\cite{bellard2005qemu}. The codes from these two datasets are highly complicated. Thus, their vulnerabilities are difficult to be identified compared to those from CWE datasets. The detailed information, including the size of training sets, validation sets, test sets, numbers of vulnerabilities, and numbers of vulnerability-free are listed in Table~\ref{table:dataset}.

\begin{figure*}[!htb]
	\centering
	\subfloat[FFmpeg Loss\label{fig:FFmpeg_Loss}]{
		\includegraphics[width=2.3in]{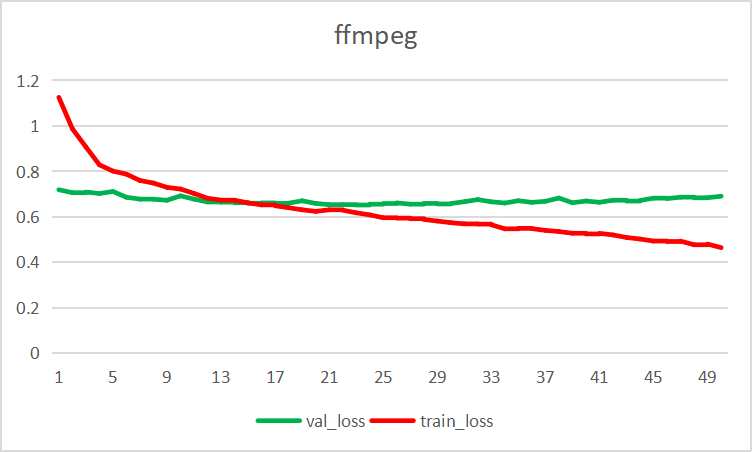}
	}
	\subfloat[QEMU Loss\label{fig:Qemu_Loss}]
	{
		\includegraphics[width=2.3in]{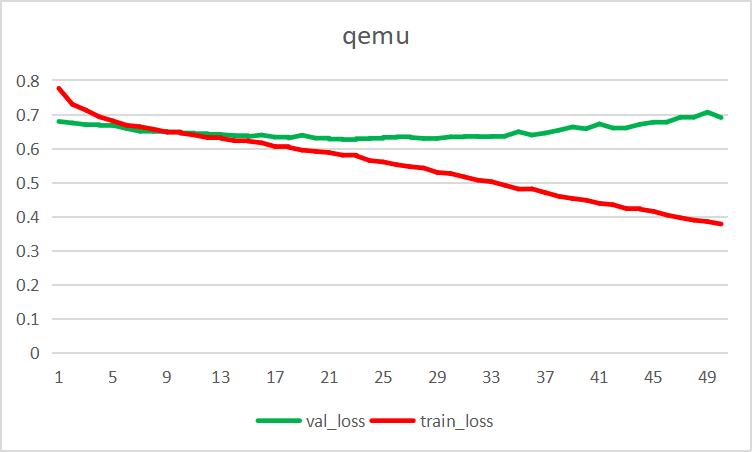}
	}
	\subfloat[CWE-362 Loss\label{fig:cwe362_Loss}]
	{
		\includegraphics[width=2.3in]{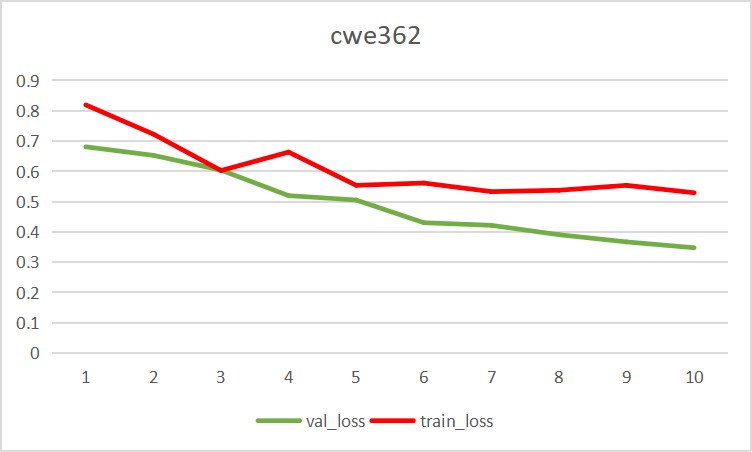}
	}

	\subfloat[CWE-476 Loss\label{fig:cwe476_Loss}]{
		\includegraphics[width=2.3in]{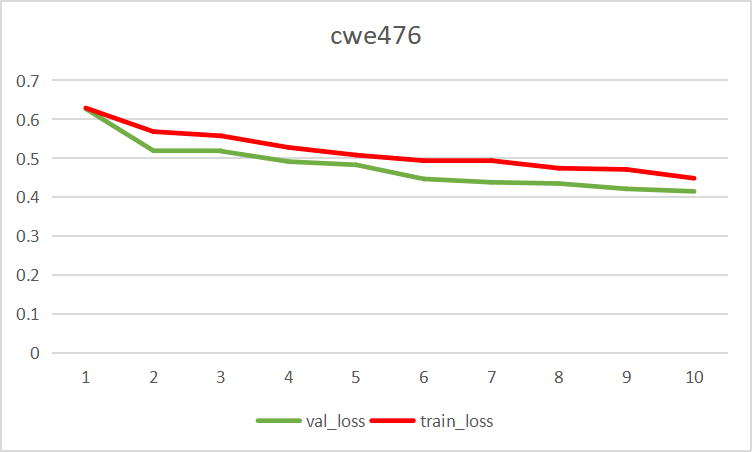}
	}
	\subfloat[CWE-754 Loss\label{fig:cwe754_Loss}]
	{
		\includegraphics[width=2.3in]{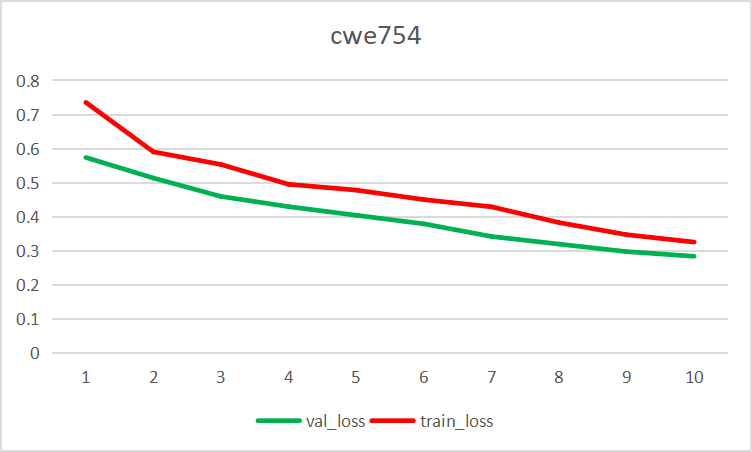}
	}
	\subfloat[CWE-758 Loss\label{fig:cwe758_Loss}]
	{
		\includegraphics[width=2.3in]{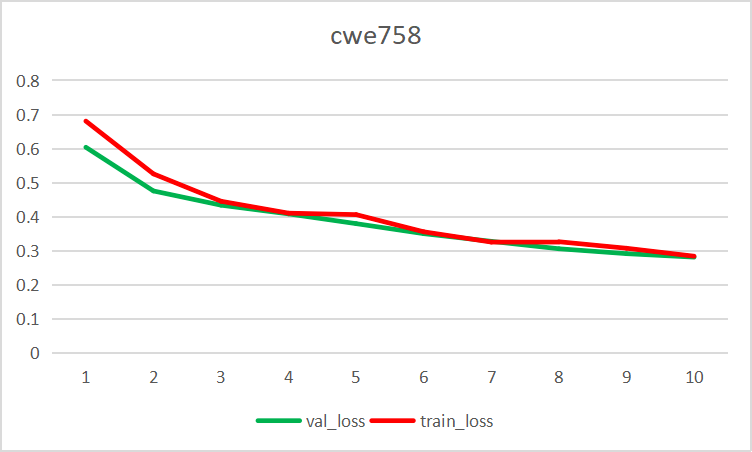}
	}
	\caption{Training Loss}
	\label{fig:loss}
\end{figure*}

\subsubsection{Environment Configuration}
The hardware configuration of our experiment is trained and tested on a multi-core computing server with a 16-core 2.10 GHz Intel Xeon CPU and an NVIDIA 3090 GPU. The RAM of the server is 256 GB and the VRAM is 24GB. The software configuration of our experiment is Tensorflow v.2.7.0 and Keras v.2.7.0. running on Windows 10. For GRSA, we set head number to 4 for PLS and 2 for CFG, DFG and AST. For the layers of convonlutional and fully-connected, we use the Adam optimizer with learning rate of 1e-5 and batch size of 32. The overall training process lasts about 1.5 hours. The final trained model has more than 300,000 hyperparameters.

\subsection{Baseline Methods}
We compared the performance of our model with six baseline methods to fully demonstrate the advantage of our model.
\subsubsection{VulDeePecker ~\cite{li2018vuldeepecker}}
 It treats the source code as natural languages and inputs the tokenized code into a multi-layer BILSTM. The initial embeddings of tokens are trained via Word2Vec.
\subsubsection{Convolution Neural Network \cite{kim-2014-convolutional}}
It treats source codes as natural languages, in which the embedding initialization is the same as that of VulDeePecker. Using  Convolution Neural Network (CNN) to extract features from source codes.
\subsubsection{SELFATT ~\cite{DBLP:journals/corr/VaswaniSPUJGKP17}}
It transfers a piece of source codes to a token sequence and exploits the multi-head attention mechanism to learn vulnerability patterns. 
\subsubsection{Devign ~\cite{zhou2019devign}}
In the composite code representation, with ASTs as the backbone, it explicitly encodes the program control and data dependency at different levels into a joint graph of heterogeneous edges with each type denoting the connection regarding the corresponding representation and utilizes the gated graph neural network model with the convolution module for graph-level classification.
\subsubsection{CodeBERT ~\cite{CodeBert}}
It is a pre-trained model for programming languages, which is a multi-programming-lingual model pre-trained on natural-language-to-programming-language (NL-PL) pairs in six different programming languages.
\subsubsection{FUNDED ~\cite{FUNDED}}
It is a graph-learning-based approach for learning code vulnerability detection models. It extends the standard graph neural network to model multiple code relationships that are essential for modeling program structures for vulnerability detection.

\subsection{Experimental Results}
We evaluate DeepVulSeeker against six baseline methods on all six datasets shown in Table~\ref{table:ffmpeg_result}. DeepVulSeeker outperforms the majorarity of baselines on almost all CWE datasets and outperforms all baselines on FFmpeg and QEMU. According to the experimental results, we summarize the following findings:

\begin{table*}[]
	\centering
	\caption{Experimental Results on CWE}
	\label{table:cwe_result}
\begin{tabular}{cccccccccccc}
	\hline
	\multirow{2}{*}{\textbf{Method}} & \multicolumn{2}{c}{\textbf{CWE-476}} &  & \multicolumn{2}{c}{\textbf{CWE-758}} &  & \multicolumn{2}{c}{\textbf{CWE-362}} & \textbf{} & \multicolumn{2}{c}{\textbf{CWE-754}} \\ \cline{2-3} \cline{5-6} \cline{8-9} \cline{11-12} 
	& ACC               & F1               &  & ACC               & F1               &  & ACC               & F1               &           & ACC               & F1               \\ \hline
	CNN~\cite{kim-2014-convolutional}                 & 0.5819            & 0.3812           &  & 0.6731            & 0.2031           &  & 0.6162            & 0.5083           &           & 0.7371            & 0.343            \\
	CodeBERT~\cite{CodeBert}    &     0.9083    &   0.8387       &  &   0.8356      &  0.7177      &  &   0.7691       &     0.8695     &           &    0.5246       &   0.4706     \\
	SELFATT~\cite{DBLP:journals/corr/VaswaniSPUJGKP17}              & 0.8401            & 0.6213           &  & 0.9121            & 0.8656           &  & 0.6627            & 0.0307           &           & 0.9219            & 0.8769           \\
	Devign~\cite{zhou2019devign}            &    0.8250      &   0.5333      &  &   0.8822      &   0.8164     &  &  0.8571      &   0.8148      &           &  0.9291      &  0.8794         \\
	VulDeepecker~\cite{li2018vuldeepecker}                &0.8070              & 0.8932         &  &  0.7887         &  0.8818      &  &  0.9500      &    0.9048 &           &    0.9574         & 0.9351            \\
	FUNDED~\cite{FUNDED}                   & 0.8889            & 0.9087           &  & 0.9583            & 0.9615           &  & 0.9446            & 0.9521           &           & 0.9286            & 0.9331           \\
	DeepVulSeeker                    & 0.9080            & 0.8052           &  & 0.9927            & 0.9859           &  & 0.9123            & 0.8387           &           & 0.9999            & 0.9999           \\ \hline
\end{tabular}
\end{table*}

\begin{table}[ht]
	\centering
	\caption{Experimental Results on FFmpeg and QEMU}
	\resizebox{\columnwidth}{!}{
	\label{table:ffmpeg_result}
\begin{tabular}{@{}cccccc@{}}
	\toprule
	\multirow{2}{*}{\textbf{Method}} & \multicolumn{2}{c}{\textbf{FFmpeg}} &  & \multicolumn{2}{c}{\textbf{QEMU}} \\ \cmidrule(l){2-6} 
	& ACC              & F1               &  & ACC             & F1              \\ \midrule
	CNN~\cite{kim-2014-convolutional}                & 0.5493           & 0.5341           &  & 0.5892          & 0.1330          \\
	VulDeePecker~\cite{li2018vuldeepecker}                &   0.5637         &  0.5702        &  &   0.5980      &   0.6141      \\
	SELFATT~\cite{DBLP:journals/corr/VaswaniSPUJGKP17}                  & 0.5710           & 0.5275           &  & 0.6049          & 0.4904          \\
	CodeBERT~\cite{CodeBert}           &  0.5264           &  0.6138           &  &  0.5354       &   0.7126       \\
	Devign~\cite{zhou2019devign}                & 0.5904           & 0.6015           &  &  0.6039        &  0.3244         \\
	FUNDED~\cite{FUNDED}              & 0.5420           & 0.6800           &  & 0.5970        & 0.7480       \\
	DeepVulSeeker                           & 0.6354           & 0.6703           &  & 0.6409          & 0.5293          \\ \bottomrule
\end{tabular}
}
\end{table}

\begin{itemize}
\item \textbf{The structural information assists the model to promote the ability of vulnerability identification.} 
Comparing CNN and VulDeePecker, we find that the Accuracy is significantly improved in both six datasets, implying that the local characteristics learned by CNN are indeed helpful for correctly identifying vulnerabilities. Also, the fact that FUNDED outperforms CodeBERT with an average of 1.56\% higher Accuracy and 6.66\% F1 Score supports the finding.

\item \textbf{The effect of attention and pre-training mechanisms are positive.}  
The Accuracy of SELFATT on FFmpeg and QEMU is 57.10\% and 57.25\% higher than that of VulDeePecker respectively. In addition, CodeBERT performs better than SELFATT across all metrics on two datasets, which fully reflects the advantages of the pre-trained model.

\item \textbf{DeepVulSeeker may be the leading approach to date.}  
On the one hand, even though DeepVulSeeker slightly underperforms some baselines on some CWE datasets in regards to either the accuracy or F1, the gap is somehow trivial to tell the difference. On the other hand, DeepVulSeeker makes a breakthrough in discovering vulnerabilities in highly-complicated datasets, yielding it more properly to be adapted to real-world industry tasks.
\item \textbf{Complicated datasets are more susceptible to the overfitting problem.} 
We show the loss changing trends during the training of our model in Fig.~\ref{fig:loss} and notice that our model has ideal loss declines in CWE dataset while less ideal loss declines in QEMU and FFmpeg data, meaning that QEMU and FFmpeg datasets are more susceptible overfitting problem. Future research can be explored on improving this problem.
\end{itemize}

\subsection{Ablation Study}
We also conduct the alation study to explore how different model components can affect the overall performance. We use FFmpeg and QEMU datasets for this study.

\textbf{Structural Information.} We first conduct ablation on the structural information of the model. The results are shown in Table.~\ref{table:ablation}. One can see that when we remove any of the graphs, it causes drops in both accuracy and F1 scores, meaning that structural information is critical for better performance. As for the FFmpeg dataset, AST and CFG play more important roles than DFG since removing either of them results in more drastic declines than removing DFG. As for the QEMU dataset, the impacts of CFG and DFG are greater than the ones of AST. Altogether, CFG may be the most significant graph among all three graphs, even though removing any of them may not be a wise option.

\textbf{Semantic Information and Pre-trained Model.} 
Applying semantic information and pre-training mechanism to the model is one of the biggest novel contributions produced by our approach. Thus, it is necessary to study the performance by removing them. In our model, we can simply remove the PLS process to achieve gain the results. As one can see in Table.~\ref{table:ablation}, after removing the PLS, the performance of the model decreases significantly on both accuracy and F1 scores. This indicates that PLS process generates and provides crucial information to our model. If these pieces of information are missing, the performance would be not much better than a random guess.

\begin{table}[htbp]
	\centering
	\caption{Ablation Study}
	\resizebox{\columnwidth}{!}{
	\label{table:ablation}
	\begin{tabular}{@{}ccccc@{}}
		\toprule
		\multirow{2}{*}{Mehods} & \multicolumn{2}{c}{FFmpeg} & \multicolumn{2}{c}{QEMU} \\ \cmidrule(l){2-5} 
		& ACC          & F1          & ACC         & F1         \\ \midrule
		DeepVulSeeker                  &\textbf{ 0.6354 }      & \textbf{0.6703  }    & \textbf{0.6409}      & \textbf{0.5293}     \\
		DeepVulSeeker w/o AST          & 0.5723       & 0.5711      & 0.6295      & 0.4816     \\
		DeepVulSeeker w/o CFG          & 0.5723       & 0.5821      & 0.6188      & 0.4405     \\
		DeepVulSeeker w/o DFG          & 0.5906       & 0.6138      & 0.6188      & 0.4056     \\
		DeepVulSeeker w/o PLS          & 0.5377       & 0.5343      & 0.5566      & 0.2422     \\ \bottomrule
	\end{tabular}
	}
\end{table}

\subsection{Case Study}
To further explore the robustness and intelligence level of DeepVulSeeker, we conduct a study on three typical vulnerable cases from the FFmpeg dataset. These three cases are correctly identified by our model but are not identified by other methods. We first analyze the vulnerability specifications of the cases. We then manually patch the vulnerability and feed them again to our model to observe whether our model would no longer flag them in order to check if our model can truly understand the meaning of the vulnerabilities in depth.

\subsubsection{Case 1} 
\begin{figure}[htbp]
	\includegraphics[width=85mm]{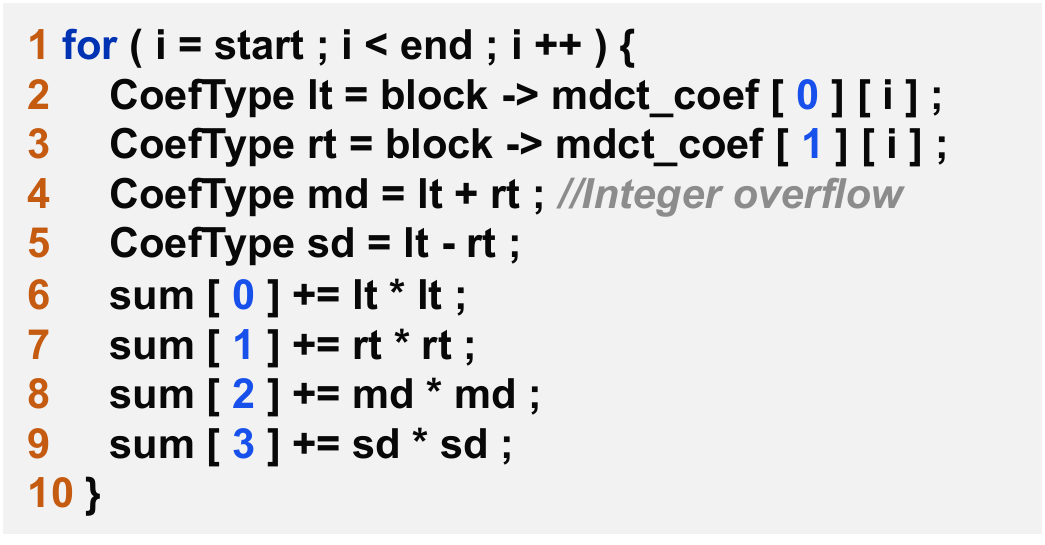}
	\centering
	\caption{Case 1}
	\label{fig:case1}
\end{figure}

In case 1 (See in Fig.~\ref{fig:case1}), the code snippet is potentially vulnerable to integer overflow. In lines 2, 3, and 4, variables \texttt{lt}, \texttt{rt}, and \texttt{md} are defined as \texttt{CoefType}, which is in fact an unsigned 32-bit integer defined in FFmpeg headers. In line 4, \texttt{md} is assigned with the sum of \texttt{lt} and \texttt{rt}, which may be greater than 32-bit, causing a possible integer overflow problem. To patch this vulnerability, we simply delete line 4. We find out that DeepVulSeeker no longer flags this code snippet after the patching, implying that it can accurately understand the meaning of the integer overflow problem.

\subsubsection{Case 2} 
\begin{figure}[htbp]
	\includegraphics[width=85mm]{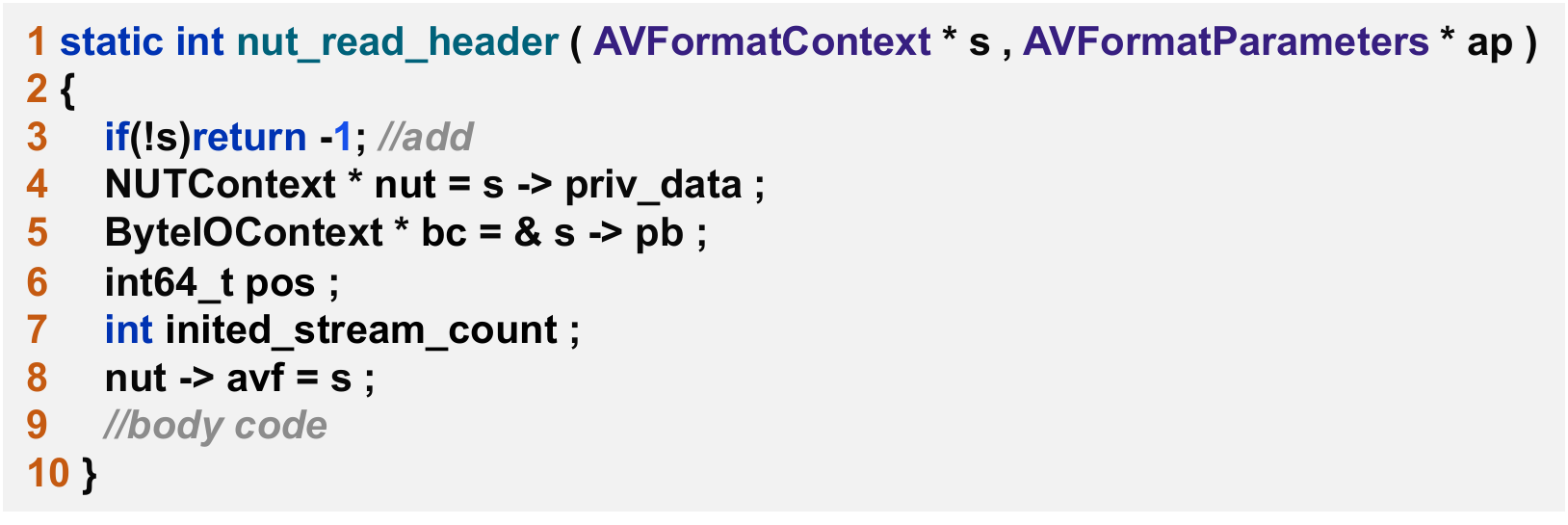}
	\centering
	\caption{Case 2}
	\label{fig:case2}
\end{figure}

In line 3 of case2 (See in Fig.~\ref{fig:case2}), the program directly accesses the pointer \texttt{s} without first checking if it is a null pointer, meaning that either the memory of \texttt{s} is freed or it is not initialized. If \texttt{s} is null, then it is a dangling pointer, which can possibly lead to serious security problems such as remote code execution (RCE) by hackers. To patch this vulnerability, we simply add a statement ``\texttt{if(!s) return -1;}'' at the beginning of the function to check if \texttt{s} is a null pointer. After the patch, DeepVulSeeker no longer reports the program as vulnerable, implying that it is able to precisely understand the reason behind the dangling pointer problem.

\subsubsection{Case 3} 
\begin{figure}[htbp]
	\includegraphics[width=85mm]{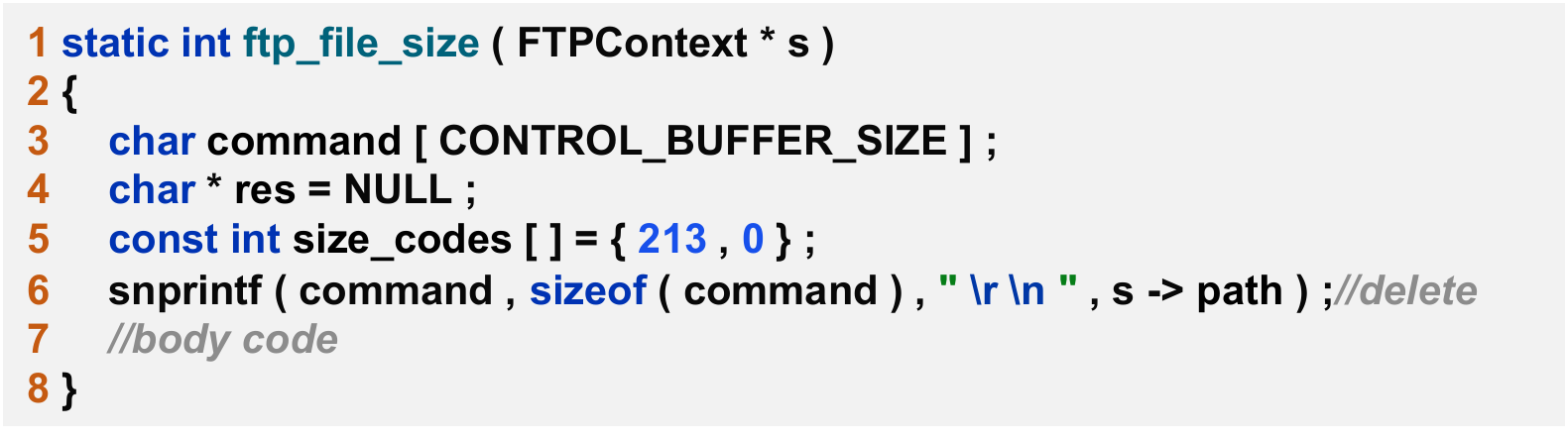}
	\centering
	\caption{Case 3}
	\label{fig:case3}
\end{figure}

In case3 (See in Fig.~\ref{fig:case3}), a possible formatted string vulnerability can happen in line 7 when invoking the function \texttt{snprintf}. The problem with this code snippet is that \texttt{snprintf} directly takes the externally-controlled format string \texttt{s->path} without regulating the format specifications before writing it to \texttt{command}. If an attacker deliberately inserts dangerous formatted specifiers such as \texttt{\%x} and \texttt{\%n} into the string, it can lead to severe consequences, e.g., memory leakage, memory corruption, or even RCE. To patch this problem, we simply delete line 7. Again, DeepVulSeeker no longer reports the program as vulnerable after the patch, indicating that it is able to learn the insights of the formatted string problem.

\section{Related Works}
\label{sec:related}

The research on software vulnerability indentification be divided into three categories, i.e., traditional detection methods, machine learning-based methods, and deep learning-based methods.

\textbf{Traditional Detection Methods.} 
Traditional detection methods refer to those that solely rely on fixed algorithms or manual efforts to discover security deficiencies, without assistance from artificial intelligence. These methods are mainly static analysis, taint-tracking, symbolic execution, fuzzing, and dynamic analysis. Wu \emph{et al.} employ static program analysis and natural language processing to differentially identify whether the vulnerability manifests a higher or lower severity in the target version~\cite{vadayath2022arbiter}. Zhang \emph{et al.} utilize taint-tracking to identify high-order vulnerabilities in OS kernels~\cite{zhang2021statically}. Kang \emph{et al.} utilize the taint analysis to extract vulnerable traces and establish a signature database for detecting recurring vulnerabilities~\cite{kang2022tracer}. Yagemann \emph{et al.} use symbolic execution to reconstruct program states and identify a couple of 0-day bugs~\cite{yagemann2021automated}. Dinh \emph{et al.} fuzzing the binding code of javascript to detect vulnerability~\cite{dinh2021favocado}. Vadayath \emph{et al.} exploit the advantages of both static and dynamic analysis to discover vulnerabilities in binary programs~\cite{vadayath2022arbiter}. Even though These methods can achieve acceptable results, they either consume great labor costs or cannot achieve full automation.

\textbf{Machine Learning-based Methods.} 
Machine learning-based methods refer to those that only employ basic machine learning methods, instead of deep learning approaches, to identify bugs. Shi \emph{et al.} use a random forest algorithm and SVM support vector machine algorithm ~\cite{he2018comparative}.  Al-Yaseen \emph{et al.} propose a multi-level hybrid intrusion detection model that uses a support vector machine and extreme learning machine to improve the efficiency of detecting known and unknown attacks~\cite{al2017multi}. Ren \emph{et al.} propose DVCMA~\cite{DVCMA}, a method for detecting software vulnerabilities based on clustering and model analyzing, and clustering technology is introduced to mine patterns from the set of vulnerability sequences in this method. Doffou \emph{et al.} propose a model~\cite{10.22937/IJCSNS.2021.21.9.46}, which is developed by combining the Multiple Correspondence Analysis (MCA), the Elbow procedure, and the K-Means Algorithm. The problem with these methods is that they may result in poor performance when dealing with complicated codes.

%Machine learning-based methods refer to those who 
%Bhoopchand \emph{et al.} represent the code as a sequence of tokens~\cite{bhoopchand2016learning}, while Raychev \emph{et al.} model the syntax tree structure of code~\cite{raychev2016probabilistic}. Peng \emph{et al.}~\cite{peng2015building} and Wang \emph{et al.}~\cite{wang2016automatically} map the nodes of abstract syntax trees extracted from programs to vectors. Allamanis\emph{et al.} use graphs to represent both the syntactic and semantic structure of code and use graph-based deep learning methods to learn to reason over program structures~\cite{allamanis2017learning}. Li \emph{et al.} model the source code as vectors for vulnerability prediction~\cite{li2018vuldeepecker}. Using machine learning is beneficial to improve accuracy, but they need to extract features or patterns hand-crafted by human experts to detect vulnerabilities.

\textbf{Deep Learning-based Methods.} 
Deep learning-based methods refer to those that leverage deep learning for identifying vulnerabilities. 
Zhou \emph{et al.}~\cite{zhou2019devign} integrate multiple source code graph structural information to obtain code semantic representation. However, they do not use the pre-training mechanism to improve accuracy and some graph node information will be lost when the graph information is integrated. Tsankov \emph{et al.} encode the dependency graph of Ethereum contracts to analyze smart contracts ~\cite{tsankov2018securify}. Nguyen \emph{et al.} represent Ethereum smart contracts as heterogeneous contract graphs to detect new vulnerabilities accurately at the line-level and contract-level~\cite{nguyen2022mando}. Liu \emph{et al.} explore combining deep learning with expert patterns for smart contract vulnerability detection~\cite{liu2021smart}. Tanwar \emph{et al.} combine the effectiveness of recurrent, convolutional, and self-attention networks towards decoding the vulnerability hotspots in code~\cite{tanwar2021multi}. Zhuang \emph{et al.} explore using graph neural networks (GNNs) for smart contract vulnerability detection~\cite{zhuang2020smart}. Dong \emph{et al.}  introduce deep-learning-based named entity recognition (NER) and relation extraction (RE) to recognize previous unseen software names and versions based on sentence structure and contexts~\cite{dong2019towards}. Feng \emph{et al.} propose CodeBERT which is a bimodal model for programming language and natural language trained by Masked Language Modeling and Replaced Token Detection~\cite{feng2020codebert}. Guo \emph{et al.} present GraphCodeBERT ~\cite{guo2020graphcodebert}, which considers the inherent structure of code by Edge Prediction and Node Alignment to support tasks like code clone detection ~\cite{huang2021code, ji2021code, sheneamer2021effective}. A variant of Roberta ~\cite{liu1907roberta} in ~\cite{wang2021codet5} supports both code understanding and generation. Alqarni
\emph{et al.} balance and fine-tune BERT~\cite{kenton2019bert} to low-level source code vulnerability detection~\cite{alqarnilow}. Compared to current deep learning-based methods, our method not only takes advantage of the structural and semantic information of a program, but also provides more effective features due to the introduction of pre-trained models.

\section{Conclusions and Future Research}
\label{sec:conclusion}
%In this paper, we propose DeepVulSeeker, a novel heterogeneous graph neural network  model, which integrates structural information including Abstract Syntax Tree, Data Flow Graph, and Control Flow Graph, to obtain multiple features of the source code for vulnerability identification. Besides, we introduce a graph autocoding attention network model to improve the performance of code representation. The experiments on six datasets demonstrate that structural information and graph autocoding attention network model are effective for vulnerability identification. In the future, we plan to introduce the knowledge graph to vulnerability identification and generate human-readable or explainable vulnerability assessment.

In this paper, we propose DeepVulSeeker, a novel model for code vulnerability identification. DeepVulSeeker is capable of acknowledging both structural information and semantic information from the codes by integrating the state-of-the-art technologies of pre-trained models, graph neural networks, and the self-attention mechanism. We test DeepVulSeeker on large heterogeneous datasets which contain conventional and highly-complicated codes, and the experimental results are more satisfactory than most of the cutting-edge baseline methods. We also conduct an ablation study and case study to further explore the model in depth. Even though our model outperforms other methods on QEMU and FFmpeg datasets, the accuracies and F1 still have rooms to be improved. We probe into the problem and find that the codes in these two datasets frequently contain inline assembly language, which cannot be handled by our method so far. In the future, we intend to revise our model with new algorithms to resolve the problem.

\section*{Acknowledgment}
This study was supported by the National Natural Science Foundation of China (62002067) and the Guangzhou Youth Talent of Science (QT20220101174).

\bibliographystyle{elsarticle-num}
\bibliography{IEEE}

\end{document}